\documentclass[12pt,aps,prd,preprint]{revtex4}
\usepackage[utf8x]{inputenc}
\usepackage{amsmath}
\usepackage{amsfonts}
\usepackage{amssymb}
\usepackage{graphicx}

\begin{document}
\title{Baryon inhomogeneities in a charged quark gluon plasma}
\author{Avijeet Ray}
 \affiliation{Indian Institute Of Technology Roorkee,  Uttarakhand, India 247667}
\author{Soma Sanyal}
\affiliation{School of Physics, University of Hyderabad, Gachibowli, Hyderabad, India 500046}

\begin{abstract}

We study the generation of baryon inhomogeneities in regions of the quark gluon plasma
which have a charge imbalance. We find that the overdensity in the baryon lumps
for positively charged particles is different from the overdensity due to the negatively
charged particles. Since quarks are charged particles, the probability of forming neutrons
or protons in the lumps would thus be changed. The probability of forming hadrons having
quarks of the same charges would be enhanced. This might have interesting consequences
for the inhomogeneous nucleosynthesis calculations. 

\end{abstract}

\maketitle

\section{Introduction}

Baryon inhomogeneities are a consequence of a first order quark hadron phase transition.
These inhomogneities formed at the QCD scale can affect the nucleosynthesis calculations.
Lattice QCD simulations indicate that the cosmic quark-hadron
transition is a crossover \cite{aoki,fodor}.  
Though there were alternative approaches which indicated that the cosmological QCD phase transition can be 
a first order phase transition \cite{agasian, tsue,ray},
current lattice studies indicate that the phase transition 
is a crossover at all times even in the presence of magnetic fields \cite{bali}. This conclusion 
from current lattice results makes baryon inhomogeneities unlikely 
in the early universe.  

However Mintz et al.\cite{mintz} show that there is a possibitlity of a first order 
phase transition when there is a finite baryon chemical potential. 
Schwarz has included the  lepton asymmetry and found that the order of the phase transition becomes
unknown for a large lepton asymmetry \cite{schwarz}.   In ref. \cite{tawfik}, the author discusses in some detail,the order of the 
phase transition for the different number of quark flavours and their masses. They mention that the order of the 
phase transition may be first, second or crossover depending on different factors in the early universe. 
Though they also discuss a crossover transition, a significant part of their discussion assumes 
a first order phase transition via bubble nucleation.  There have been several studies on quark nuggets \cite{lugones} and their 
consequences in the early universe. As of now, both baryon inhomogeneities and quark nuggets are formed only if the 
phase transition is a first order transition in the early universe. As mentioned in ref. \cite{florkowski}, the probability 
of formation of such objects is small but they are not completely ruled out. Recently Boeckel et al. \cite{boeckel} came up 
with a novel idea that opens up the possibility of a first order QCD phase transition in the early universe. 
This was followed by several papers which considered the consequences of a strong first order phase transition in the 
early universe \cite{caprini}. They were trying to obtain observational signatures of the QCD phase transition using the new 
gravitational wave detectors. Since Boeckel's idea has reopened the possibility of a first order QCD phase transition \cite{simon} 
and as baryon inhomogeneities are a natural consequence of a first order phase transition \cite{tawfik, layek, mocsy}, we would like
to study the generation of baryon inhomogeneities in regions of the quark gluon plasma
which have a charge imbalance.

Observations indicate that our universe is charge neutral over large lengthscales.
However, there is the possibility of generating charge separated domains over smaller
lengthscales at different epochs during the evolution of the universe. Such domains can
be generated during reheating after the cosmological inflation \cite{gasenzer}, or from
the lepton asymmetry after the electroweak scale \cite{zarembo}. Others, \cite{mahato,bigazzi} 
have also referred to the charge asymmetry of the early universe plasma and its 
consequences. Since we are looking at large regions with a charge or anti charge overdensity, 
the charged plasma in ref. \cite{gasenzer} and  ref.\cite{zarembo} would be the starting point for us.  
As a large lepton asymmetry is not excluded experimentally  Zarembo showed that the quark gluon plasma
above the transition temperature can be electrically charged. Gasenzer et al. obtained distinct 
domains of charge and anti-charge overdensities after the cosmological inflation.  
It is possible that these charge separated regions persist over parts of the universe upto the QCD epoch.
In ref.\cite{dodin}, it is shown that charge oscillations are damped by the expansion of the universe,
a numerical analysis of the equations at relevent temperatures  
will show that the oscillations survive over a timescale that is large compared to timescales
in the early universe. However, since we are not interested in obtaining the size of the charged domains, 
we contend that the charge imbalance may have decreased due to the expansion of the universe, but it has not been
wiped out completely. The plasma will then be a charged quark gluon plasma similar to that 
proposed by Zarembo \cite{zarembo}.  Consequently, we would like to investigate the effect of such regions of charged quark
gluon plasma on the formation of baryon inhomogeneities.

As it is a first order phase transition, we are interested in charge
separation over a distance greater than the thickness of a  bubble wall. 
Bubbles are nucleated with a radius which varies in size from $1 fm - 10^3 fm$ depending
on the supercooling. Thin wall bubbles have
a bubble wall thickness of about $1 fm$ \cite{sigl}. So we are interested in charge
separation scenarios which will generate patches of charge imbalance with sizes greater
than  $1 fm$. This charge separation gives
rise to an electrostatic potential. The presence of this potential affects the transport
of the charged quarks through the bubble wall. We show
that this leads to a change in the amplitude of the baryon inhomogeneities generated and 
also allows for the possibility of forming charged baryon inhomogeneities which have not 
been discussed in the literature before. 

There is also the possibility that if the phase transition is a spinodal decomposition
with the formation of domains of true and false vacuums, the charge fluctuations can
affect the nature of domain formation which may lead to exotic candidates for
dark matter. We have not explored this in detail in this paper but hope to pursue it
later.

\section{Effect of charge separation on baryon inhomogeneities}

 A first order phase transition 
takes place with the nucleation of hadronic bubbles in a quark gluon plasma phase. This is followed by a period of co-existence of the two 
phases. Since the thermodynamic potentials have different dependencies on the chemical potential $\mu$, a baryon number density contrast 
builds up on the two sides of the phase boundary. Detailed calculations show that these
baryon inhomogeneities may affect the nucleosynthesis 
calculations  \cite{layek2,ssanyal,olesen,fuller}. We are interested to know whether the presence
of an electric potential in the quark gluon plasma affects the density 
contrast of the inhomogeneities. In the presence of a non-zero charge density,we can 
obtain a potential $\phi$. We consider regions in the quark gluon plasma which have a
charge imbalance and are of sizes larger than $1 fm$. If ($n_{+} - n_{-}$) is the
net charge overdensity in such a region then, the potential can be obtained by solving
Poisson's equation.
\begin{equation}
 \nabla^2 \phi(x) = 4 \pi e (n_{-}-n_{+})
\end{equation}
Since the plasma is in thermodynamic equilibrium, we may expect,
\begin{equation}
  n_{-}-n_{+} = n_0 \left [e^{( \frac {-e \phi}{ T} )} - e^{( \frac {e \phi}{ T} )}
\right]
\end{equation}
We can get an approximate value for the potential by using the Debye screening model,
\begin {equation}
 \phi \approx \frac {e^{-(\frac {r}{\lambda_d})}}{r}
\end {equation}
where $r = \sqrt{x^2+y^2+z^2}$ and $ \lambda_d = (\frac{T}{8 \pi n e^2})^{1/2} $ is the
Debye length.
$ \lambda_d \approx 1 fm $ during the quark- hadron transition.
The Debye length sets the smallest lengthscale in the plasma.
This indicates that for thin walled bubbles, the bubble wall thickness is of the 
same order of magnitude as the Debye length. It is possible to obtain an approximate value
of the potential  by considering a charge density 
 of the order of $10^{2} MeV^3 $  instead of a point charge \cite{gasenzer}. We obtain  $\phi \sim
 0.1 MeV$. This is an approximate value but it tells us that the potential generated can
be in the MeV range.

In the presence of this potential, the grand canonical function of the QGP phase is given
by \cite{sigl}, 
\begin{equation}
 \frac{\Omega_{qgp}}{V} = -\Sigma \frac{21 \pi^2 T^4}{180} \left [1 + \frac {30}{7 \pi^2} \left (\frac {\mu_i - q_i \phi}{T} \right )^2 \right] + \frac {E^2}{8 \pi} 
\end{equation}
where $\mu_i $ and  $q_i $ are the chemical potential and the electric charge of the particular species of particle. 
We  consider $\Omega$ in both the phases to be of the lowest order. The baryon number density in both the phases can 
be obtained from their respective grand canonical functions. 

For the QGP phase where there are regions of localized electric fields, the number density
of the $i^{th}$ particle is given by, 
\begin{equation}
 n_i^q \sim \frac{9}{27} T^3 \left (\frac{\mu_{qi} - q_i \phi}{T} \right )
\end{equation}

The newly nucleated hadronic phase does not have any charge imbalance  and hence the grand
canonical function in this phase is given by,
\begin{equation}
 \frac{\Omega_{had}}{V} = \frac{-2gT^4}{\pi^2} \Sigma \frac{{(-1)}^{n+1}}{n^4} cosh \left
[\frac {n \mu_{hi}}{T} \right] {\bar {K_2}}(nm/T)
\end{equation}
 
For $\frac{\mu_b}{T} \sim 10^{-8}$, the number density is given by,
\begin{equation}
 n_i^h \sim \left [\frac{8}{\pi^3} \right ]^{1/2} T^3 \left (\frac{\mu_{hi}}{T} \right )
\left (\frac{m}{T} \right )^{3/2} e^{\frac{-m}{T}}
\end{equation}

In the limit of chemical equilibrium across the bubble wall, the baryon chemical potential can be taken to be the same for both the phases. 
Therefore, $\mu_{qi} = \mu_{hi} = \mu_{b} $. Then the baryon overdensity in the hadronic phase is given by, 
\begin{equation}
 R = \frac{n^q}{n^h} = R_0 \left (\frac{\mu_b - q \phi}{\mu_b} \right )
\end{equation}
where $R_0 = \frac{2}{9} \left[{\frac{\pi^3}{8}} \right]^{1/2} \left[\frac{T_c}{m}\right
]^{3/2} e^{m/T_c}$ is the density contrast in equilibrium in the absence of an electric
potential \cite{fuller}.

This expression shows that the number density 
of the negatively charged particles goes up while the number density for the positively
charged particles decreases. Due to the simplified model used, the
density contrast increases or decreases linearly compared to its value in equilibrium.
Moreover, the value of $\frac {\mu_b}{T} \approx 10^{-8}$, hence even for
small values of $\phi$, the contrast deviates significantly from the equilibrium value.

We give an example of this deviation in fig 1. The solid line gives the  density contrast
($R_0$) in equilibrium in the absence of an electric potential. The  dashed line gives the
overdensity for the negatively charged particles ($q = -1$). We have used a log scale on
the Y-axis,
to show how the density contrast increases with potential. As the potential
increases the density contrast deviates rapidly. Thus it is clear that negatively and
positively charged particles will be selectively filtered by the hadronic bubble wall.

We have used a very simplified model to demonstrate the possibility of filtering
charged particles through bubble walls. This expression of $R$ does not take into
account the transport probabilities across the hadronic bubble wall. It will work only in 
the limit of very rapid transportation of baryon number across the phase boundary. We now proceed to do a more detailed 
calculation which involves the net flux of quarks incident on the hadronic bubble wall and includes a term which filters the charged particles 
depending on the value of the potential. We show that even in this case the baryon
inhomogeneity  $R$ is different for quarks of opposite charges.  

\begin{figure}
\includegraphics[width = 62mm,angle = 270]{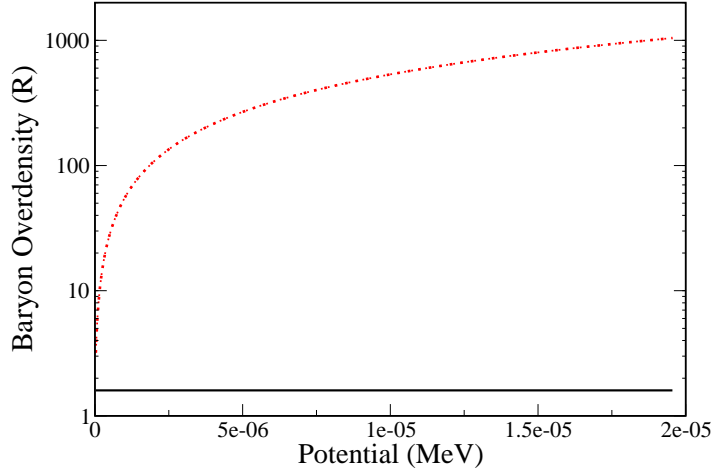}
\caption{Baryon overdensity of negatively charged particles varying with the potential.
The dashed line gives the
overdensity for the negatively charged particles in presence of a potential, the solid 
line gives the density contrast $R_0$, in the absence of a potential  }      
\end{figure} 
 
\section{Baryon number transport across the hadronic bubble wall}

Baryon transport across the bubble wall has been discussed previuosly in the absence of a
potential \cite{fuller}. We now do the
detailed calculations factoring in the presence of the potential.  
We first consider the transport of baryon number from the unconfined to the confined
phase. The presence of the potential leads to the separation of charges 
on the bubble wall. The charge density will therefore fall of exponentially close to the
bubble wall. We denote the charge density in this region by $n_{\pm}$ and it can be
represented by 
\begin{equation}
 n_{\pm} = n_0 e^{\mp \frac { \phi} {3 T}}
\end{equation}
Here $n_0$ is the equilibrium number density in the respective phase. For simplicity, the
equilibrium number density is considered to be the same for all the quarks. We will
denote $n_{\pm}^q $  for number density of quarks in the QGP phase and  $n_{\pm}^h $
for the number density in the hadronic phase.
Assuming that the QGP side of the wall has a net negative charge while the hadronic side
has the compensatory positive charge, we write the
total flux of the quarks as a summation of the positive and negative particles. 
\begin{equation}
 f = n_0 \left [e^{( \frac {- \phi}{3 T} )} + e^{( \frac { \phi}{3 T} )}
\right] 
\end{equation}
The total recombination rate per unit area will depend upon the probability of combining three quarks of the right color and flavor. However due to the 
presence of the potential, quarks of some flavors will dominate over quarks of other flavors. Hence the probability of forming hadrons with quarks of 
both charges are decreased. For the time being we do not distinguish between the hadrons but only mention that the probability of forming neutrons 
and protons would be less than the probability of forming hadrons like $\Delta^{++}$, $\Omega^{-}$ etc. So the recombination rate will depend upon the flux
and the cube of the quark number density. The total recombination rate per unit area of the 
boundary is then given by,
\begin{equation}
 \Lambda_q = (1.3 \times 10^{-5}) f \Sigma_q (\frac {T}{100 MeV})^9 \chi \chi'  
\end{equation}
Here $\Sigma_q$ is the dimensionless transmission probability for recombination, $\chi$ is
the net number of quarks over antiquarks divided by the total number of quarks of all
kinds and
$\chi'$ is the net number of positively charged particles over negatively charged particles divided by the total number of quarks of all kinds. 
The quark-antiquark factor is given by,
\begin{equation}
 \chi \approx 0.61 \mu_b /T   
\end{equation}
The factor $\chi'$ arises due to the fact that more positive charges will move towards the bubble wall than negative charges. Expanding the number density 
as a Taylor series and ignoring the higher order terms, we get 
\begin{equation}
 \chi' \approx - {\frac{\phi}  {3 \mu_b}}   
\end{equation}
Unlike $\chi$, the sign on $\chi'$ will depend on the nature of the potential and the
charge of the quark moving towards the bubble wall. This will lead to selective
filtration of charged quarks into the hadronic bubble. 
The filter factor $F$ is obtained by dividing the net number of baryons encountered by
the front per unit area in time $\Delta t$ by the net number recombined 
at the front per unit area in time  $\Delta t$.  Therefore, 
\begin{equation}
F_{\pm} \equiv \frac{\Delta N_b} {N_b}   
\end{equation}
where, 
\begin{equation}
N_b (cm^{-2})=   n_0 \left [e^{( \frac {- \phi}{3 T} )} - e^{( \frac { \phi}{3 T} )} \right] (V_f) \left [\frac{\Delta t}{10^{-6}} \right],   
\end{equation}
$V_f$ is the velocity of the front and $\Delta N_b (cm^{-2})= \Lambda_q \Delta t $. 
Hence 
\begin{equation}
F_{\pm} = \pm 0.8 \times 10^{-13} \left [ \frac{\phi} {3 V_f} \right] \Sigma_q \left[
\frac{T} {100} \right ]^8 
\end{equation}

We have been calculating the baryon number transfer rate across the front
from the QGP to the hadronic phase. There is also
the possibility of reverse baryon number transport from the hadronic phase to the QGP. At the constant temperature $T_c$, the system will move towards 
chemical equilibrium where the chemical potential on both sides of the boundary are equal and the baryon transfer rate per unit area in both phases 
become equal. We use this equilibrium condition to derive the baryon number transport equations in time. 

Assuming a low $\mu_b$ in the hadronic phase the flux of baryon number moving towards the wall is given by,      
\begin{equation}
 f_h = \left[\frac{8}{3 \pi^3}\right]^{1/2} c T^2 m \left[\frac{\mu_b}{T} \right] e^{-m/T} \left [e^{( \frac {- \phi}{3 T} )} - e^{( \frac { \phi}{3 T} )} \right] 
\end{equation}
Here we have taken the nucleons to be moving with the speed of light $c$. If the probability that a hadron gets through the phase boundary is given by 
$\Sigma_h$ then the rate of baryon number transfer per unit area from the hadron phase to the quark gluon plasma phase is, 
\begin{equation}
 \Lambda_h \equiv f_h \Sigma_h
\end{equation}

At chemical equilibrium $\Lambda_h = \Lambda_q $ and therefore $\Sigma_h$ can be related to  $\Sigma_q$.
\begin{equation}
 \Sigma_h  = \left [(8.8 \times 10^{-6})  \left (\frac {T}{100 MeV} \right )^{10} \left
(\frac{\phi}{T} \right ) e^{(\frac{-m}{T})} \right]  \Sigma_q
\end{equation}
The baryon transport across the bubble wall takes place during the constant temperature epoch when the two phases coexist in thermal equilibrium.
If we denote the volume fraction of the universe in the QGP phase to be $V_{q}$. Then as the phase transition proceeds $V_q$ decreases and the volume fraction 
in the hadronic phase increases. Apart from this the universe is also expanding. The total baryon number per comoving volume is then given by 
 a constant $N_0$ such that,  
\begin{equation}
 \frac{N_0}{V} = (n_+^q + n_-^q) V_{q} +  (n_+^h + n_-^h) (1-V_{q})
\end{equation}
Here $(n_+^q + n_-^q)$ gives the baryon number corresponding to the positively and negatively charged particles in the QGP phase and $(n_+^h + n_-^h)$ are the
corresponding values in the hadronic phase.
The detailed dynamics of the growing bubbles in the context of an expanding universe has
been dealt with in several papers before \cite{fuller}. Using the Robertson 
-Walker metric, the expansion of the universe is related to the radius of the expanding bubbles as well as the fractional change in volume due to the phase 
transition. The rate of change of volume due to the expansion of the universe is given by 
\begin{equation}
 \frac{\dot V}{V} = \frac{3 \dot A }{A}
\end{equation}
If $N_n$ be the number of bubbles nucleated within the horizon and $r$ be the typical
radius of the nucleated bubbles, then the total baryon number that passes 
through the bubble wall is given by $n_b^q \lambda_{\pm}^q $, where $\lambda_{\pm}^q$ is
the baryon
transfer rates from the QGP to the hadronic phase for the particles of different charges.
The time evolution of the baryon number will also depend
on the charge of the baryon. Therefore, the transfer rates from the QGP to the hadronic
phase become,
\begin{equation}
 n_{\pm}^q \lambda_{\pm}^q = 4 \pi r^2 N_n \frac{V_0 }{V f_v} V_f F_{\pm}
\end{equation}
where the fraction of volume in the QGP phase is given by $f_v$, the transfer rates are
obtained as
\begin{equation}
  \lambda_{\pm}^q =  \frac{V_f F_{\pm}} {n_{\pm}^q}
\end{equation}
Similarly one can obtain the baryon transfer rates from the hadronic to the QGP phase. 
The transmission probability of the baryon number fron the hadronic phase to the QGP
phase is given by $\Sigma_h$. We find out the number of hadrons hitting the total surface
area of the bubble walls. Since we have already specified the number of bubbles nucleated
in a given horizon volume $V_0$ as $N_n$, the total surface area of the nucleated bubbles
becomes $4 \pi r^2 N_n$.  
The baryon transfer rate is then given by, 

\begin{equation}
\lambda_{\pm}^h = \frac{1}{3} \left[\frac{4 \pi r^2 N_n V_0}{f_v V}\right] V_b \Sigma_h 
\end{equation}
Once the baryon transfer rates are known for all the charged particles in the two phases
one can obtain the time evolution of the baryon number in the respective phases,
\begin{equation}
 \dot{n_{\pm}^q} = -n_{\pm}^q \lambda_{\pm}^q + n_{\pm}^h \lambda_{\pm}^h -n_{\pm}^q \left
[\frac{\dot{V}}{V} + \frac{\dot{V_f}}{V_f} \right]
\end{equation}
\begin{equation}
 \dot{n_{\pm}^h} = \left[\frac{V_f}{(1-V_f)} \right]\left[-n_{\pm}^h \lambda_{\pm}^h +
n_{\pm}^q \lambda_{\pm}^q -n_{\pm}^h \left (\frac{\dot{V_f}}{V_f} \right)
\right] - n_{\pm}^h \frac{\dot{V}}{V}
\end{equation}

The only free parameters are the transmission probabilities through the phase boundaries.
Since these two are related to each other through eqn (34), hence we have only one free
parameter to deal with. As discussed previously by Fuller et al. \cite{fuller}the value
of the transmission probability will  depend on the model chosen to define the phase
boundary between the confined and the unconfined phases.  We keep it as a free
parameter and obtain the overdensities based on a range of $\Sigma_h$ values.
We find that the overdensity is different for positive and negative particles but the
difference does not depend on the  $\Sigma_h$ value. As expected, it depends on the
potential. Fig.2 shows the ratio of the overdensity for positive and negative particles
for a potential of $50 MeV$.  
 \begin{figure}
\includegraphics[width = 66mm]{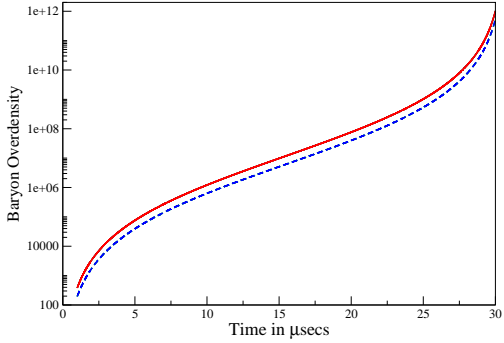}
\caption{Baryon overdensity for charged particles in the presence of a potential. The 
dashed line gives the
overdensity for the negatively charged particles, the solid 
line gives the density contrast for the positively charged particles.} 
 
\end{figure} 
The figure indicates that the number of positively charged particles in the baryon
overdensities would be more than the negatively charged particles. So the probability of
forming positively charged hadrons increases in comparison to the other hadrons. This
will affect the neutron proton ratio at the time of nucleosynthesis. It may also lead to
the formation of positively charged baryon overdensities. Charged baryon inhomogeneities
and their consequences have not been explored so far. It is quite possible that the decay
of such inhomnogeneities during a later epoch may address some of the questions about
unknown radiation sources that have been observed lately. 

All the calculations here are done with a positive potential. If we use a negative
potential, we see that the overdensity is more for negative particles. So the charge
asymmetry in the inhomogenities depends on the sign and value of the potential. Since the
overall charge of the plasma remains neutral, we will get both positive as well as
negative potential regions. Hence at the end of the quark hadron transition, we will be
left with charged as well as neutral baryon inhomogenities. More detailed simulations
will be able to acertain whether the charged inhomogeneities will be negligible in
comparison to the neutral ones. 

\section{Conclusions}

In this work, we have looked at baryon number transport through a bubble wall in patches
of quark gluon plasma where there is a charge imbalance. If there are such regions, they
would have some local potential which would affect the baryon transport rate across the
bubble wall. This baryon transport rate is primarily responsible for the generation
of baryon inhomogeneities in the early universe. We have obtained an expression for the
baryon contrast in presence of such an electromagnetic potential. We find that 
baryon overdensities are enhanced depending on the charges of the particles. 
This seems to indicate that the presence of the electric potential results in regions
of charged overdensities. These charged overdensities will decay over time. 
Since the decay of baryon inhomnogeneities cause a change in the neutron proton ratio,
they will affect the nucleosynthesis results. It is also possible that  selective
filtering of charged particles through the bubble
walls may lead to more stable objects which will decay at a much later 
epoch. The decay of these objects might explain several unexplained sources of energy 
detected in the present universe.

\begin{center}
 Acknowledgments
  
\end{center}
 
We would like to thank Rajarshi Ray for his comments on this work.

\end{document}